# A Horserace of Volatility Models for Cryptocurrency: Evidence from Bitcoin Spot and Option Markets*

Yeguang Chi[1], Wenyan Hao[2]

September 2020

**Abstract**

We test various volatility models using the Bitcoin spot price series. Our models include HIST, EMA, ARCH, GARCH, and EGARCH models. Both of our in-sample-fit and out-of-sample-forecast results suggest that GARCH and EGARCH models perform much better than the other models. Moreover, the EGARCH model's asymmetric term is positive and insignificant, which suggests that Bitcoin prices lack the asymmetric volatility response to past returns. Finally, we formulate an option trading strategy by exploiting the volatility spread between the GARCH volatility forecast and the option's implied volatility. We show that a simple volatility-spread trading strategy with delta-hedging can yield robust profits.

**Keywords:** *volatility estimation, volatility forecasting, cryptocurrency trading, option pricing*

* We thank the helpful comments from Lianxin He, James Zhang, Leo Tao, and Leo Guo. We also thank the data support from Grandline Technologies®. All errors are our own.

[1] Auckland University Graduate School of Management, 12 Grafton Road Sir Owen G Glenn Building, Auckland, New Zealand 1010. Email: y.chi@auckland.ac.nz.
[2] University of Leicester, Department of Mathematics, University Road, Leicester, United Kingdom, LE1 7RH. Email: wenyan881224@gmail.com.

# 1. Introduction

Since Nakamoto (2008)'s seminal paper on Bitcoin, cryptocurrencies have taken both the technological and investing worlds by storm. Led by Bitcoin, cryptocurrencies have been commonly associated with not only their astronomical price appreciations in their early developmental years, but also their mind-boggling volatilities that have caused many to doubt their true potential to replace fiat currencies. Despite a lot of controversies, the fact that cryptocurrencies have become an emerging asset class to many investors worldwide is indisputable. It is therefore important to understand the volatilities associated with this new asset class. Coupled with the fact that cryptocurrency options are also becoming more widely available to investors, it makes the task of volatility estimation and forecasting even more relevant at the current time.

In our paper, we take on the challenge of modeling Bitcoin's volatility. In particular, we conduct a horserace among several most commonly used volatility models: historical average (HIST), exponentially-weighted moving average (EMA), autoregressive conditional heteroscedasticity (ARCH), generalized autoregressive conditional heteroscedasticity (GARCH), and exponential generalized autoregressive conditional heteroscedasticity (EGARCH) models. Our comparative statistics regarding these models' in-sample fit and out-of-sample forecasting power clearly suggest the winners to be GARCH and EGARCH models. Moreover, we observe no asymmetric volatility response to past returns in the Bitcoin price series. Furthermore, we contrast the GARCH model's volatility forecast with the option implied volatility, and again confirms the GARCH model's superior performance in forecasting future realized volatilities. Finally, we apply our GARCH model forecast to formulate a simple volatilty-spread trading strategy that shows robust profits.

Our paper contributes to several finance literatures. First, we add to the empirical literature that investigates volatility models to fit and forecast financial time series. This line of research is rich with both financial series and volatility models (e.g. Engle 1982, Bollerslev, 1986, Andersen et al., 2003, Dyhrerg, 2016, Katsiampa, 2017, Hu et al., 2019). Specifically, Andersen and Bollerslev (1998) argue that standard volatility models do provide accurate forecasts. Brooks (2014) offers a comprehensive discussion regarding commonly used standard volatility models. Among them, we pick the HIST, EMA, ARCH, GARCH, EGARCH models for our study using Bitcoin price series. Consistent with many studies before us, we find that the parsimonious GARCH model is hard to beat.

Second, we provide new evidence on the asymmetric volatility response to past returns using the Bitcoin price series. Since Black (1976)'s study of stock price volatility changes, the relationship between volatility to and returns has been documented by many studies using various financial price series. For example, French, Schwert, and Stambaugh (1987) find that unexpected stock market returns are negatively related to the unexpected changes in the volatility of stock returns; Campbell and Hentschel (1992) modify the GARCH model of stock market returns to allow for an asymmetric volatility feedback effect that amplifies large negative stock returns and dampens large positive returns; Wang and Wang (2011) document asymmetric volatilities in the Chinese stock market and attribute it to unconfirmed asymmetry in return reversals; Carr and Wu (2017) examine the S&P 500 index returns and identify a self-exciting behavior associated with large negative market disruptions, which helps explain the asymmetrically increasing market volatility after negative market returns. To address the asymmetric volatility response, econometricians have developed a variety of modified GARCH model. A common version is the EGARCH model (e.g. Nelson, 1991, McAleer and Hafner 2014). We employ the EGARCH model in our investigation of the Bitcoin price series and show that the asymmetric term is positive and statistically insignificant. In other words, we do not observe the common stock market phenomenon of increasing volatility following negative returns with the Bitcoin price series.

Third, we investigate a novel dataset on cryptocurrency options from the Deribit exchange. Various studies on modeling financial series' volatility has used options data (e.g. Bartunek, 1995, Blair et al. 2001, Alexander et al., 2020a, 2020b, Alexander and Imeraj, 2020). The cryptocurrency options exchanges developed much later than the cryptocurrency spot and futures exchanges. Consequently, the data on cryptocurrency options is just starting to become abundant. We use the options trading data from the largest cryptocurrency options exchange Deribit® for our study. In particular, we find that option implied volatility adds little to the GARCH model in forecasting future realized volatilities. Furthermore, we contrast our GARCH model's volatility forecast with the option implied volatility to arrive at a volatility spread. We show that a simple volatility-spread trading strategy with delta hedging can provide robust profits. In other words, during our sample period, there may have existed exploitable mis-pricings on the Deribit options exchange.

The rest of our paper is organized as follows. Section 2 discusses data and methodology. Section 3 conducts the horserace among various volatility models and provide comparative statistics on their in-sample fit and out-of-sample forecasting power. Section 4 further contrasts our GARCH volatility forecast with the option implied volatility, and evaluates the performance of a simple volatility-spread trading strategy. Section 5 concludes and offers research extension. Appendix includes variable definitions and trading-cost discussions.

## 2. Data and methodology

2.1. Data source

Alexander and Dakos (2020) have offered their critique regarding the data quality in the cryptocurrency literature. Following their findings, we choose to use high-quality cryptocurrency trading data from two large and reputable cryptocurrency exchanges. Our first data source, Binance®, was launched in July 2017 and since then has developed into one of the world's largest cryptocurrency exchanges. Binance is the most liquid exchange in terms of daily trading volume as documented by Hougan (2019). For our study, we investigate the Bitcoin spot instrument on Binance (ticker: BTC.USDT), whose total trading volume amounted to 99 billion USDT[3] in 2019 (Binance, 2019). Our BTC.USDT price series is sampled at the minute-level frequency. Our sample period is from August 17, 2017 to March 26, 2020.

Our second data source, Deribit®, was launched in June 2016 and is now the largest cryptocurrency options exchange in the world. It offers plain vanilla European options on Bitcoin and Ethereum, as well as futures instruments and perpetual swaps[4]. According to Deribit statistics, the total trading volume of its option contracts is $9.16 billion for the year of 2019, far more than its competitors such as OKEx and the CME Group. For our study, we investigate the Bitcoin perpetual swaps (ticker: BTC.USD) and Bitcoin options. Our BTC.USD price series is sampled at the minute-level frequency. Our Bitcoin options trade data are at the tick level. Our sample period is from August 17, 2018 to March 26, 2020.

In section 3, we use the Binance BTC.USDT price series for the purpose of model estimation and forecasting. In section 4, we use the Deribit BTC.USD price series for the purpose of model estimation and forecasting. The BTC.USDT and BTC.USD price series are very highly correlated (correlation > 0.99). In section 3, we use the Binance data for the volatility model horserace because our Binance data is one year longer than our Deribit data. In section 4, we use the Deribit BTC.USD series for certain volatility model estimation and forecasting because BTC.USD is the underlying instrument upon which option strikes are based. Finally, for Deribit options data, we use a tick-trade dataset that includes the time, price and volume of each trade.

---

[3] USDT is a type of stable coin offered on Binance whose value is pegged to USD with a 1:1 exchange ratio.
[4] The Deribit Bitcoin perpetual swaps (also known as Bitcoin perpetual) is a derivative product similar to a future, however, without an expiry date. It employs a concept known as 'funding rate' to keep the Bitcoin perpetual swap price close to the underlying Bitcoin index price. As a result, the Bitcoin perpetual swap can be considered as a proxy to Bitcoin spot. Interested readers can find out more at: https://www.deribit.com/pages/docs/perpetual

2.2. Volatility model description

We investigate five different volatility models in our paper: historical average (HIST), exponentially-weighted moving average (EMA), autoregressive conditional heteroscedasticity (ARCH), generalized autoregressive conditional heteroscedasticity (GARCH), and exponential generalized autoregressive conditional heteroscedasticity (EGARCH) models. We now define each model following Enders (2004) below.

2.2.1. HIST model

The historical realized average volatility is defined by

$$RV_t = \sqrt{\frac{1}{N-1} \sum_{j=t-N}^{t} (r_j - \frac{1}{N} \sum_{j=t-N}^{t} r_j)^2}, \text{ where } r_t = \frac{S_t - S_{t-1}}{S_{t-1}}, \quad (1)$$

where $S_t$ marks the trading instrument's price at period $t$. Unless defined otherwise, we report annualized volatilities.

2.2.2. EMA model

Let $r_j$ be the return series, where $j \in \{1, \dots, t\}$, then $z_t^2$ is the estimate of the variance of period $t$, which is

$$z_t^2 = \lambda z_{t-1}^2 + (1 - \lambda) r_t^2, \quad (2)$$

where $r_t^2$ is the most recent return estimated over the observations and $z_{t-1}^2$ is the previous estimate of the volatility. Here, the value of $\lambda$ is $\frac{2}{(n+1)}$, where $n$ is set to 365 in our paper.

### 2.2.3. ARCH model

The ARCH model was developed by Engle (1982). It is commonly employed in modelling financial time series that describes the variance of the current error term as a function of the actual sizes of the previous time periods' error terms. Specifically, let $(v_t)_{t \in \mathbb{Z}}$ be standard normal distribution (0, 1). The process $(r_t)_{t \in \mathbb{Z}}$ is an ARCH(p) process if it is strictly stationary and if it satisfies, for all $t \in \mathbb{Z}$ and some strictly positive-valued process $(z_t)_{t \in \mathbb{Z}}$, the equations

$$r_t = z_t v_t, \quad z_t^2 = a_0 + \sum_{i=1}^{p} \alpha_i r_{t-i}^2. \tag{3}$$

where $a_0 > 0, \alpha_i \geq 0, i = 1, \dots, p$. For our study, adopt an ARCH(1) model. We refer to the ARCH(1) model simply as the ARCH model from now on.

### 2.2.4. GARCH model

The GARCH model was developed by Bollerslev (1986). This model allows the conditional variance to be dependent upon previous own lags. Specifically, let $(v_t)_{t \in \mathbb{Z}}$ be standard normal distribution (0,1). The process $(r_t)_{t \in \mathbb{Z}}$ is an GARCH(p, q) process if it is strictly stationary and if it satisfies, for all $t \in \mathbb{Z}$ and some strictly positive-valued process $(z_t)_{t \in \mathbb{Z}}$, the equations

$$r_t = z_t v_t, \quad z_t^2 = a_0 + \sum_{i=1}^{p} \alpha_i r_{t-i}^2 + \sum_{j=1}^{q} \beta_j z_{t-j}^2. \tag{4}$$

where $a_0 > 0, \alpha_i \geq 0, i = 1, \dots, p,$ and $\beta_j \geq 0, j = 1, \dots, q$. For our study, we adopt a GARCH(1, 1) model. We refer to the GARCH(1, 1) model simply as the GARCH model from now on.

2.2.5. EGARCH model

The exponential generalized autoregressive conditional heteroskedastic (EGARCH) model by Nelson (1991) is another form of the GARCH model. Empirically, it has been observed for many financial time series that the conditional volatility of returns often reacts asymmetrically to the past return shocks. The EGARCH(p, q, o) model captures the asymmetric volatility response. The conditional variance is defined by

$$\ln z_t^2 = a_0 + \sum_{i=1}^{p} \alpha_i \left(\frac{|r_{t-i}^2|}{\sqrt{z_{t-i}^2}} - \sqrt{2/\pi}\right) + \sum_{j=1}^{o} \theta_i \frac{r_{t-j}^2}{\sqrt{z_{t-j}^2}} + \sum_{k=1}^{q} \beta_k \ln z_{t-k}^2. \quad (5)$$

For our study, we adopt an EGARCH(1, 1, 1) model. We refer to the EGARCH(1, 1, 1) model simply as the EGARCH model from now on.

## 3. Horserace of volatility models

In this section, we conduct a horserace among the five volatility models: HIST, EMA, ARCH, GARCH, and EGARCH. First, we offer summary statistics on the BTC.USDT price series and discuss relevant volatility patterns. Second, we offer discussions regarding these five models' in-sample fit. Third, we compare these five models' out-of-sample forecasting power.

3.1. Summary statistics

Figure 1 plots the daily close, returns and volatility series of our spot instrument BTC.USDT. We calculate and report the annualized volatility by first computing the minute-level return standard deviation and then scaling it up to the annual frequency. From Figure 1a, we observe a wide range of Bitcoin price. From the start of our sample period in August 2017, the Bitcoin price had experienced an astronomical appreciation from sub-$5,000 level to nearly $20,000 level, within just a few months. Since then, the Bitcoin price had quickly retreated back to sub-$10,000 within a month of the price peak, before it slowly depreciated to just above $3,000 in December 2018. The first few months of 2019 were quiet with the Bitcoin price below $4,000. Starting from April 2019, the price took off in April from $4,000 all the way up to $14,000 in June

2019. Since then, the Bitcoin price has come down and eventually reached an ebb of sub-$4,000 level in March 2020. Clearly, Bitcoin price has exhibited tremendous ups and downs within our sample period.

This observation is corroborated by Figure 1b of the daily return series and by Figure 1c of the daily volatility (annualized) series. We see in Figure 1b multiple instances where daily returns reached magnitudes above 20%, with the most recent instance that took place during the March 2020 selloff with a single-day return of negative 39.5%. We see in Figure 1c that annualized volatility of the Bitcoin price series is high overall, with massive volatility spikes over 500% during at least two episodes, first of which was the bullish period during late 2017, and second of which was the bearish period during March 2020. Moreover, in Figure 1b, we observe patterns of volatility clustering, which we will study in more details in the remainder of our paper.

In Table 1, we report the various moments of the Bitcoin price series at frequencies from daily to 1-hour. Corroborating volatility patterns we observe in Figure 1c, we show that the annualized volatility is 85.83%, based on daily price series. It is interesting to see a generally positive relationship between annualized volatility and sampling frequency. For example, at the hourly (high) sampling frequency, we would reach an annualized volatility of 97.23%, much higher than the 85.83% at the daily (low) sampling frequency. We believe that this pattern is caused by the large intra-day volatility of the Bitcoin price series. That is, by looking at the price series more frequently, we are able to pick up more ups and downs in intraday price movements. We observe that skewness on the daily level is negative, which suggests that Bitcoin prices tend to be experiencing more negative outliers in its distribution than positive outliers. But the same cannot be said for other sampling frequencies. For example, we actually see positive skewness at the 12-hour, 3-hour and 1-hour sampling frequencies. Last but not least, we observe consistently large excess kurtosis at all sampling frequencies. In fact, the higher our sampling frequency, the larger the kurtosis. For example, at the 1-hour frequency, we observe an excess kurtosis of 29.74. This pattern shows that the Bitcoin price exhibits strong fat-tail property against the normal distribution, especially at higher sampling frequencies.

3.2. In-sample fit

In Table 2, we report results regarding the in-sample fit of ARCH, GARCH and EGARCH models. Our sample period is from August 17, 2017 to March 26, 2020, which includes a total of 953 daily prices. Across all three models, we observe a statistically significant coefficient $\alpha$ ($\alpha$=0.20, t=3.19 under ARCH; $\alpha$=0.11, t=2.38 under GARCH; $\alpha$=0.08, t=2.17 under EGARCH). This confirms the volatility-clustering pattern in Figure 1b. Similar to many other financial assets, large changes in Bitcoin prices tend to follow large changes and small changes tend to follow small changes. Hence, a positive $\alpha$ captures this auto-correlating feature of Bitcoin's volatility time series.

Across the GARCH and EGARCH models, we observe a significantly positive coefficient $\beta$ ($\beta$=0.83, t=17.45 under GARCH; $\beta$=0.80, t=13.80 under EGARCH). This shows that the extra lagged variance term in the GARCH and EGARCH models relative to the ARCH model plays an important role in volatility series' in-sample fitting estimation. Judged by the higher log-likelihood and lower AIC and BIC values, we can further confirm that the GARCH and EGARCH models work better than the ARCH model in fitting Bitcoin's volatility series. On the other hand, GARCH and EGARCH performance is similar to each other, with EGARCH showing slightly better statistics than GARCH.

Moreover, the extra term $\theta$ in the EGARCH model relative to the GARCH model is positive and statistically insignificant (0.13, t=1.12). Under the EGARCH model, if the negative asymmetric volatility response to past returns exists, we would anticipate a negative $\theta$. The positive and insignificant $\theta$ in our model suggests a weak and positive asymmetric volatility response to past return from Bitcoin's daily price series. That is, Bitcoin's volatility tends to rise (fall) after positive (negative) past returns. Our result using Bitcoin prices stands in contrast to those observed using stock prices (e.g. S&P500), where a significantly negative $\theta$ is observed (French, Schwert and Stambaugh, 1987; Campbell and Hentschel, 1992). One possible interpretation is that Bitcoin is not subject to the same leverage effect (Black, 1976) that stocks are subject to. Hence, Bitcoin does not necessarily have to exhibit the same asymmetric volatility response to past returns as stocks do.

Last but not least, our GARCH model parameters are well specified in the sense that our intercept term $a_0$ is positive and $(\alpha + \beta)$ is less than one. The unconditional variance implied by the GARCH parameters is $a_0/(1 - \alpha - \beta)$. After scaling it up to annualized frequency and taking the square root, we arrive at an unconditional annualized volatility estimate of 90.96%, which is quite close to our in-sample volatility estimate of 85.83% based on the daily price data.

3.3. Out-of-sample forecast

After our in-sample estimation of the GARCH model parameters, we further carry out the task of volatility forecasting. In addition to using the whole history as our look-back window, we also use various look-back windows from 30 days to 365 days to estimate the GARCH model parameters. In Table 3, we report the 1-day-ahead volatility forecast of these GARCH models with various look-back windows. In our sample, a longer look-back window produces a higher average and median volatility forecast value. For example, the average 1-day-ahead annualized volatility forecast value based on the whole historic look-back window is 71.5%, compared against 61.8% based on the 30-day look-back window; the median 1-day-ahead annualized volatility forecast value based on the whole historic look-back window is 68.4%, compared against 56.6% based on the 30-day look-back window. This finding echoes the volatility-clustering pattern shown in Figure 1b. It seems the shorter look-back windows tend to capture the more-often low-volatility periods which yield average lower 1-day-ahead volatility forecast value. However, a shorter look-back window does produce a higher maximum volatility forecast value. For example, the maximum 1-day-ahead annualized volatility forecast value based on the 30-day look-back window is 234.0%, compared against 123.4% based on the whole historic look-back window. This finding echoes the extreme volatility-spike pattern shown in Figure 1c. During those extreme volatility bursts, the shorter look-back windows tend to capture more of the short-term spikes in volatility than the longer look-back windows.

Next, we conduct the following predictive OLS regression to compare across various volatility models' out-of-sample forecasting power:

$$\text{RV}_{t+1} = \beta_0 + \beta_1 \text{FV}_t + e_{t+1} \tag{6}$$

where $\text{RV}_{t+1}$ is the future realized volatility of day $t+1$ calculated from minute-level BTC.USDT return series; $\text{FV}_t$ is the model-specific volatility forecast for day $t+1$ based on data available by the end of day $t$. We consider the whole historic look-back window, and various other look-back windows from past 365 days to past 30 days, for each model we consider.

In Table 4, we report the regression outcomes including the intercept $\beta_0$, the predictive coefficient $\beta_1$, and the adjusted R-square. Moreover, we calculate the mean absolute error (MAE) as an additional performance proxy defined as follows:

$$\text{MAE} = \frac{1}{N}\sum_{n=1}^{N} |\Delta - y_n| \tag{7}$$

where $y_n$ is the regression's model prediction and $\Delta$ is the true value.

After comparing the out-of-sample forecasting performance across these models, it is clear that GARCH and EGARCH models deliver substantially better volatility forecasts. For example, using the past 365 days as the look-back window, the GARCH forecast exhibits an adjusted R-square of 49.02%, whereas the EGARCH forecast exhibits an adjusted R-square of 46.24%, both of which are far superior to the three other models, e.g. HIST model's –0.15%, EMA model's 0.04%, and ARCH model's 10.78%. The MAE proxy corroborates the same intuition, as the GARCH forecast again shows the best (smallest) MAE value of 18.53%, whereas the EGARCH forecast shows the second best MAE value of 19.92%.

It is interesting to note that for both the HIST and EMA models, the shorter look-back windows tend to generate better out-of-sample forecasts as evidenced by higher adjusted R-squares and lower MAE. For example, the HIST model based on the past-30-day look-back window exhibits an adjusted R-square of 35.84%, far higher than 3.60% that is based on the past-180-day look-back window; the EMA model based on the past-30-day look-back window exhibits an adjusted R-square of 35.32%, far higher than 15.02% that is based on the past-180-day look-back window. Given the volatility-clustering patterns of the Bitcoin price series, such results are to be expected. This is because longer look-back windows tend to converge to the long-term historic average value of the volatility, which does not help capture the short-term clusters in volatility.

Finally, we observe mixed evidence on the GARCH model's superiority over the EGARCH model in forecasting volatility across various look-back windows. One may argue for the preference of one model over the other. We prefer to use the GARCH model over the EGARCH model for our volatility forecast application for two related reasons: first, the GARCH model is more parsimonious than its EGARCH counterpart; second, the asymmetric volatility response is statistically insignificant in our Bitcoin price series, which weakens the motivation for the EGARCH model.

We plot the 1-day-ahead volatility forecast values of the GARCH model in Figure 2 and those of various other models (HIST, EMA, ARCH, and EGARCH) in Figure 3. The dashed lines in these figures represent the 1-day-ahead future realized volatility. The solid line represents the model forecast values of the 1-day-ahead volatility. All models are based on the whole historic look-back window. Consistent with results we find in Table 4, we see that the GARCH model captures much more of the time-series variation in future realized volatility than the HIST, EMA and ARCH models; and it performs on par with the EGARCH model. The ARCH model picks up more of the time-series variation in volatility than HIST and EMA models, but fares poorly when compared against the GARCH and EGARCH models. However, there still remains a large amount of residual volatility that is not captured by the GARCH model. We see many volatility spikes in future realized volatility that still significantly deviate from the GARCH forecast. Towards the very end of our sample period in March 2020, Bitcoin has experienced a sharp selloff that has seen its volatility shoot up to nearly 600% whereas our GARCH forecast remains below 300%.

## 4. Volatility-spread trading strategy

In the last section, we have shown that the GARCH model delivers the best performance regarding both the in-sample fit and out-of-sample forecast. In this section, we apply the GARCH model's volatility forecast to Deribit data and test a simple volatility-spread trading strategy. First, we offer summary statistics on Deribit's BTC option contracts. Second, we test various volatility models' forecasting power using Deribit's BTC.USD price series, along with the option implied volatility as an additional forecasting variable. Third, we construct a volatility-spread trading strategy by comparing our GARCH volatility forecast against the option implied volatility, and evaluate its performance.

4.1. Summary statistics of Deribit BTC option contracts

In Table 5, we report the quarterly summary statistics regarding the BTC option contracts that are traded on the Deribit exchange. Since the beginning of our dataset in the fourth quarter of 2018, Deribit BTC option contracts have seen persistent growth, reflected by an increasing trend in the number of contracts offered on the exchange, the total number of trades, and the total trading volume in USD. We define the trading volume as the option premium in USD multiplied by the number of option contracts traded. In the fourth quarter of 2018, there were only 236 listed BTC option contracts associated with a total of 34,587 trades that resulted in $382 million in trading volume; by the first quarter of 2020, there were 971 listed BTC option contracts associated with a total of 121,041 trades that resulted in $3.33 billion in trading volume. Over the course of one year and half, the Deribit exchange has seen its trading volume almost grow ten-fold. As of 2020, Deribit is without doubt the most vibrant marketplace for option trading within the cryptocurrency space.

For the exercises in this section relevant to Deribit exchange's options that mature on March 27, 2020, we focus on the period during which the options are listed, i.e. from September 10, 2019 to March 26, 2020[5]. For each day $t$ from September 10, 2019 onwards, we estimate the GARCH model with the whole historic daily data of BTC.USD index price quoted from Deribit from August 17, 2018 to day $t$. We then generate the GARCH model's volatility forecast of day $t+1, t+2, ..., T$, where $T$ marks the option's maturity date. We call the volatility forecast on day $t+1$ our single-day volatility forecast. Next, we calculate the equal-weighted average of the volatility-forecast series from day $t+1$ to $T$ to arrive at the GARCH model's multi-day average volatility forecast.

On the other hand, we calculate the volume-weighted average implied volatility of at-the-money call and put options with maturity on March 27, 2020, which we refer to as option implied volatility. To determine the at-the-money strike on a given day $t$, we use the closest strike (by $1,000 interval) to the underlying Bitcoin index's value weighted average price (vwap) on day $t$. For example, if day $t$'s vwap is $7,100, we pick $7,000 as the at-the-money strike. For another example, if the vwap is $8,800, we pick $9,000 as the at-the-money strike. Once we determine the strike, we then aggregate all trades ($i=1,2,...,I$) of calls and puts at the chosen strike on day $t$ and produce the volume-weighted average $IV_t$ as follows.

$$IV_t = \frac{\sum_{i=1}^{I} w_{i,t} \, IV_{i,t}}{\sum_{i=1}^{I} w_{i,t}} \qquad (8)$$

---

[5] The actual maturity date of the option is March 27, 2020, but the last day's data is incomplete as the option matures during the day. Therefore, we choose to end our sample period on March 26, 2020.

where $IV_{i,t}$ is the option implied volatility of trade $i$, $w_{i,t}$ is the dollar trading volume for trade $i$, and $I$ is the total number of trades for day $t$.

In Figure 4, for our sample period from September 10, 2019 to March 26, 2020, we plot the option implied volatility for day $t+1$ (solid line), the GARCH single-day volatility forecast for day $t+1$ (dotted line), and the GARCH multi-day average volatility forecast for the period between day $t+1$ and day $T$ (dashed line). It is clear that the single-day volatility forecast fits the option implied volatility much worse than the multi-day average volatility forecast. The comparative result is not surprising, because the option implied volatility reflects the volatility that applies to the period until option maturity, not the volatility for just the next day. The same intuition also helps explain the fact that option implied volatility does not exhibit as much time-series variation as the GARCH single-day volatility forecast.

4.2. Option implied volatility's forecasting power

Using Deribit BTC.USD price series, we re-conduct our regression analysis to test various models' out-of-sample forecasting power on the volatility of the BTC.USD prices. In addition to the ARCH, GARCH and EGARCH models we have already considered, we also consider the Bitcoin option implied volatility as a forecasting variable. Our data from Deribit is from August 17, 2018 to March 26, 2020. For the exercises in this section relevant to Deribit exchange's quarterly options that mature on 27 March, 2020, we focus on the period during which the options are listed, i.e. from September 10, 2019 to March 26, 2020.

Specifically, for each day $t$ from September 10, 2019 onwards, we estimate each model with the whole historic daily data from the start of our Deribit data (i.e. August 17, 2018) up to day $t$ of BTC.USD index price quoted from Deribit. We then generate the model-specific volatility forecast of day $t+1$ ($ARCH_t$, $GARCH_t$, and $EGARCH_t$). Additionally, we calculate day $t$'s volume-weighted average implied volatility ($IV_t$) of at-the-money call and put options with maturity on March 27, 2020. Finally, we calculate the future realized volatility of day $t+1$ ($RV_{t+1}$) from the minute-level BTC.USD index prices and then scale it up to the annual frequency.

We investigate the volatility-forecasting power of various model combinations based on the following regression:

$$RV_{t+1} = \beta_0 + \beta_1 * ARCH_t + \beta_2 * GARCH_t + \beta_3 * EGARCH_t + \beta_4 * IV_t + e_{t+1}. \qquad (9)$$

In Table 6, we report the predictive regression's outcomes of each single model, as well as combinations of the previously specified model and option implied volatility. Similar to what we observe in Table 4, the models that lead the pack are still the GARCH and EGARCH models. Option implied volatility does not forecast future realized volatility well, as its performance is similar to that of ARCH but far worse than the GARCH and EGARCH models. For example, the adjusted R-square for the IV model is 31.83%, compared with the GARCH model's 63.83% and the EGARCH model's 64.92%. Moreover, none of the model combinations generate any meaningful improvement over the GARCH model. For example, the addition of IV to either the GARCH or the EGARCH model yields little improvement, if any.

4.3. Volatility-spread trading strategy

In this section, we develop a simple volatility-spread trading strategy. The volatility spread is defined as the difference (former minus latter) between our GARCH model's multi-period average volatility forecast (referred to as GARCH forecast below) and the given option's implied volatility. We consider the GARCH model with sampling frequency of 24 hours (i.e. daily) that estimates the GARCH model using daily data and updates volatility forecast every 24 hours. In unreported results, we also try the 12-hour sampling frequency in addition to the 24-hour sampling frequency and find similar results. We will provide results upon request to interested readers.

Our trading instrument for the backtest is the $8000-strike call option with maturity on March 27, 2020 (BTC8000C27MAR20). We choose this contract for two reasons: first, it is a relatively new contract so that it has better liquidity than earlier contracts; second, its strike $8,000 is close to the median price level of the BTC.USD index during the sample period, which ensures that it is one of the most liquid option instruments. The option started trading on September 10, 2019 and expired on March 27, 2020. We update the option implied volatility each time there is a trade in our dataset. At every update, we compute the volatility spread between our latest GARCH forecast and the option implied volatility.

Two key parameters governing our trading strategy are the volatility spread's entry threshold and exit threshold. Given an entry threshold, we would enter the trade as soon as the volatility spread's magnitude crosses the threshold. For example, assuming an entry threshold of 0.05, if the volatility spread of the current trade is 0.06 and the volatility spread of the last trade is 0.04, then we would enter a 1-unit long position on the call option and simultaneously short delta[6] unit of the underlying (i.e. Bitcoin perpetual swap). For another example, assuming an entry threshold of 0.10, if the volatility spread of the current trade is –0.12 and the volatility spread of the last trade is –0.08, then we would enter a 1-unit short position on the call option and simultaneously long delta unit of the underlying.

After our trade entry, we would then hold the position and implement standard delta-hedging procedure every time the call option's delta changes by more than 0.02. Delta-hedging would keep our total position delta-neutral so that we can focus on the PNL solely generated from the volatility spread. We would not exit the position until the exit threshold of the volatility spread is reached. For example, assuming an exit threshold of 0.00 and our original position is long call option and short underlying, if the volatility spread of the current trade is –0.01 and the volatility spread of the last trade is 0.02, then we would exit our position by selling the call option and buying back the underlying. For another example, assuming an exit threshold of 0.05 and our original position is long call option and short underlying, if the volatility spread of the current trade is –0.01 and the volatility spread of the last trade is 0.02, then we would keep holding our position because we would wait until the volatility spread crosses –0.05 to exit our position. A positive exit threshold essentially means that we would wait until the volatility spread sufficiently crosses to the other side of zero to exit our position. Continuing the last example, if the volatility spread of the current trade is –0.06 and the volatility spread of the last trade is –0.04, then we would exit the position by selling the call option and buying back the underlying.

In Table 7, we report the backtest results (including number of trades, win/loss ratio, win rate, total PNL and PNL per trade) for various specifications of entry threshold and exit threshold. Overall, results are encouraging. Across all specifications, we observe high win rates, large win/loss ratios and positive PNLs. For example, the (24-hour/0.05/0.00) specification generates 25 trades with a 88.5% win rate and a total PNL of $2,251. Moreover, we estimate our initial capital cost to be $5,000, which should be sufficient in meeting both the option premium and portfolio margin requirement of our strategy[7]. We discuss trading costs on Deribit exchange in more detail in Appendix A2.

---

[6] 'delta' stands for the option delta as inferred by standard Black-Scholes option-pricing formula (Black and Scholes, 1973).

In Figures 5, we plot the PNL curve of the volatility-spread strategy based on the GARCH model with entry parameter set at 0.05 and exit parameter set at 0.00. We also mark the trade entries in black dots in the figure. Interested readers may match these trades with a more detailed trade summary presented in Tables 8. It is interesting to observe that our strategy has registered many unprofitable trades in March 2020, due to the massive spikes in volatility that took place mid-March. This was the time when the Bitcoin price tanked 50% within less than a week. It may be argued that the volatility regime has switched during this time, which caused our volatility-spread strategy to stop working, as the strategy relies on historic data and may not be able to react fast enough to the volatility regime switch. How to identify regime switch and react accordingly in a volatility trading strategy is a meaningful research question. We leave such investigations to future research.

4.4. Trading strategy incorporating volatility smile

In the previous subsection, we compare the GARCH model volatility forecast with the implied volatility of the $8000-strike call option with maturity on March 27, 2020 (BTC8000C27MAR20). On the one hand, the GARCH model is fitted using the underlying BTC.USD instrument, which means its volatility forecast is that of the underlying (i.e. at-the-money price level). On the other hand, the BTC.USD price fluctuates and does not constantly stay around the $8000 level. Therefore, it would be more precise if we consider the difference between implied volatilities at different moneyness levels, also known as volatility smile (Hull, 2003). In this subsection, we incorporate the effect of volatility smile (i.e. at-the-money options exhibit lower implied volatility than in-the-money and out-of-the-money options) into our GARCH volatility forecast. Specifically, we follow the steps below.

Step 1: pick one day each month from October 2019 to March 2020 on which the BTC.USD close price is near a thousand-dollar multiple (e.g. $7,000, $8,000, $9,000 or

---

[7] For more information regarding the margin requirement, one may consult the following link: *https://www.deribit.com/pages/docs/portfoliomargin*.

$10,000). Dates we picked include November 01, 2019, December 06, 2019, January 07, 2020, February 04, 2020, and March 05, 2020).

Step 2: for each date we picked, we estimate the volatility smile's slope by using the two 1000-dollar adjacent call options. For example, on March 05, 2020, the BTC.USD closed at around $9,000 level, so we define the BTC8000C27MAR20 contract and BTC10000C27MAR20 as the two adjacent contracts. We calculate the implied volatilities of the three contracts BTC8000C27MAR20, BTC9000C27MAR20, and BTC10000C27MAR20 to be $IV_{8000}$, $IV_{9000}$, and $IV_{10000}$. We also calculate the option deltas of these three options to be $\delta_{8000}$, $\delta_{9000}$, and $\delta_{10000}$. Next, we calculate the average slope of the volatility smile on March 05, 2020 as $(\frac{IV_{8000}-IV_{9000}}{|\delta_{8000}-\delta_{9000}|} + \frac{IV_{10000}-IV_{9000}}{|\delta_{9000}-\delta_{10000}|})/2$. Finally, we take an equal-weighted average across the five dates we picked to arrive at the average slope $s_{avg}$ of value 0.150. We assume 5% risk-free rate, and 70%[8] volatility for these calculations.

Step 3: by the end of each trading day *t* in our sample, calculate the adjustment term for our GARCH volatility forecast as $s_{avg} \times |\delta_{8000}^t - \delta_{close}^t|$, where $\delta_{8000}^t$ marks the option delta of the trading instrument BTC8000C27MAR20, and $\delta_{close}^t$ marks the option delta calculated for a call option with strike exactly at day *t*'s close price, and maturity on March 27, 2020. Consequently, our volatility forecast for day *t+1* now becomes the sum of our GARCH forecast and the adjustment term $s_{avg} \times |\delta_{8000}^t - \delta_{close}^t|$.

Step 4: implement the same volatility-spread trading strategy, except that we now use the new volatility forecast calculated in Step 3.

We plot the cumulative PNL, and trade entries for the trading strategy in Figure 6, similar to Figure 5. We observe similar trading performance for the specification (24hour/0.05/0.00), whether we control for the volatility smile or not. We also report the trading performance for various other trading specifications in Table 9. We observe mixed evidence on trading performance after we incorporate the volatility smile effect. For example, in the (24hour/0.075/0.05) specification, the total PNL has improved from $1,553 to $2,178; the PNL per trade has improved from $155 to $182; but the win rate has decreased from 90.0% to 77.8%. For another example, in the (24hour/0.05/0.00) specification, the total PNL has decreased from $2,251 to $2,028; but the PNL per trade has improved from $90 to $97; and the win rate has improved from 88.5% to 95.2%.

---

[8] We use 70% volatility assumption for this exercise because it is the average volatility during our sample period from September 11, 2019 to March 27, 2020.

## 5. Conclusion and extension

In this paper, we have conducted a horserace among five volatility models (HIST, EMA, ARCH, GARCH, EGARCH) using Bitcoin price series obtained from Binance. We find consistent evidence that GARCH and EGARCH models outperform the rest of the pack in terms of both in-sample-fit and out-of-sample-forecast performance. Moreover, Bitcoin prices do not exhibit asymmetric volatility response to past returns, a pattern commonly observed with stock price data.

Next, we apply the GARCH volatility forecast to the Deribit data and find that it is better at forecasting future realized volatility than option implied volatility. We then formulate an option trading strategy that exploits the volatility spread between the GARCH volatility forecast and the option implied volatility. We show that a simple volatility-spread trading strategy with delta-hedging can yield robust profits, even after considering trading costs.

Currently, we utilize just one price series (i.e. Bitcoin) to estimate our GARCH volatility model. In future research, we believe it may be fruitful to explore other cryptocurrencies' price series under the framework of a multi-GARCH model. It would be interesting to explore the potential benefits of interacting multiple price series in volatility estimation and forecasting. Last but not least, as Deribit offers both Bitcoin and Ethereum options, it also remains to be seen how a multi-GARCH model that incorporates both BTC and ETH price series may compete with a standalone GARCH model.

# Appendix

## Appendix A1: Variable definitions

| Variables | Definition & Description |
|---|---|
| $RV_t$ | Realized volatility, calculated as the standard deviation of BTC returns. In the context of the HIST model, we calculate the historical realized volatility in daily frequency during different lookback windows, such as past 365 trading days, 180 trading days, 90 trading days, and 30 trading days. In the context of volatility forecasting, we calculate the future realized volatility in minute frequency of day *t+1* and then scale it up to the annual frequency. |
| $FV_t$ | Model-forecasted volatility, calculated in daily frequency by different models including the EMA model, ARCH model, GARCH model, and EGARCH model |
| $IV_t$ | Option implied volatility, calculated by fitting the Black-Scholes option pricing formula's analytic solution to the option premium. We assume an annual risk-free rate of 5% for our calculation. |

**Appendix A2: Trading cost assumptions**

Deribit has a maker-taker fee model. You become a 'maker' when you place an order, and it does not trade immediately, so your order stays in the order book and waits for someone else to fill/match with it later. The 'taker' is someone who decides to place an order that is instantly matched with an existing order on the orderbook.

Assuming that we fill our orders as 'takers' on the Deribit exchange, we need to pay 0.04% of the underlying instrument (or 0.0004 BTC) per options contract. Moreover, option fees can never exceed 12.5% of the option premium. For BTC perpetual swaps (the Bitcoin underlying instrument), the taker fee is 0.075%. For more details on trading fees of various Deribit contracts, interested readers can consult the following link: *https://www.deribit.com/pages/information/fees*. Throughout the course of this trade from September 10, 2019 to March 26, 2020, we would incur option trading costs of $85 on 25 trades, and delta-hedging costs of $361 on a total trading volume of 59.46 BTC perpetual swaps. The total trading costs of $446 constitutes about 19.8% of our total PNL of $2,251.

Finally, we also consider the bid-ask spread of both the option contract and the underlying Bitcoin perpetual swap instrument. For option contracts that are at or near the money, the bid-ask spread is on average around 3% of the option premium. For Bitcoin perpetual swap instrument, the bid-ask spread is normally only $0.50. With the conservative cost assumption that we always meet opponent prices in executing orders, our trading costs in the form of bid-ask spread amount to $1,149 for the option part and $33 for the BTC perpetual swaps. The total market impact of $1,182 constitutes about 52.5 % of our total PNL of $2,251.

## Declaration of Interest

We do not have any conflict of interests to declare.

## Funding Acknowledgement

This research did not receive any specific grant from funding agencies in the public, commercial, or not-for-profit sectors. We thank Grandline Technologies® for data support.

**Figure 1**
**Daily series of BTC.USDT's price, return and volatility**

Figures 1a, 1b, and 1c plot BTC.USDT's daily close, returns and volatility (annualized), respectively. In calculating annualized volatility of a given day, we use the day's minute-level return data to calculate a minute-level volatility which we then scale to annual frequency. Our sample period is from August 17, 2017 to March 26, 2020.

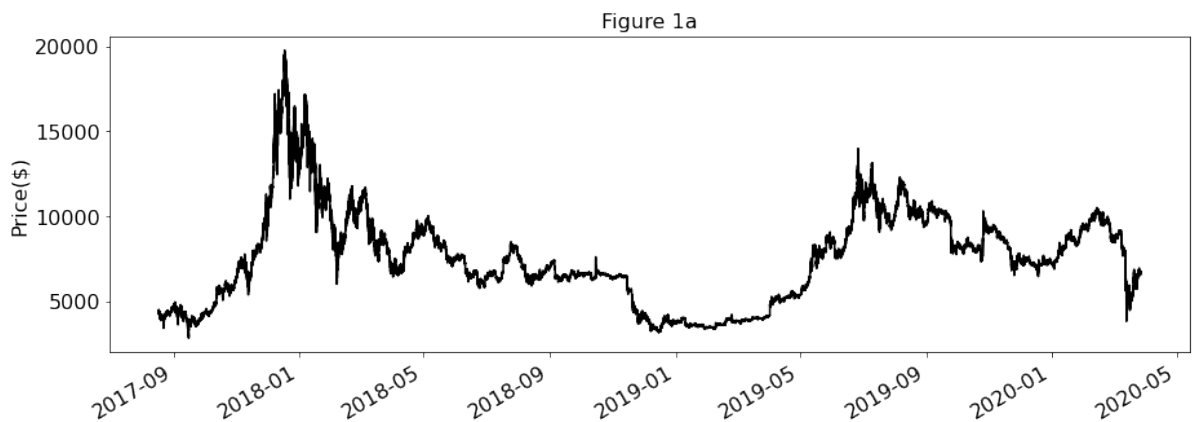

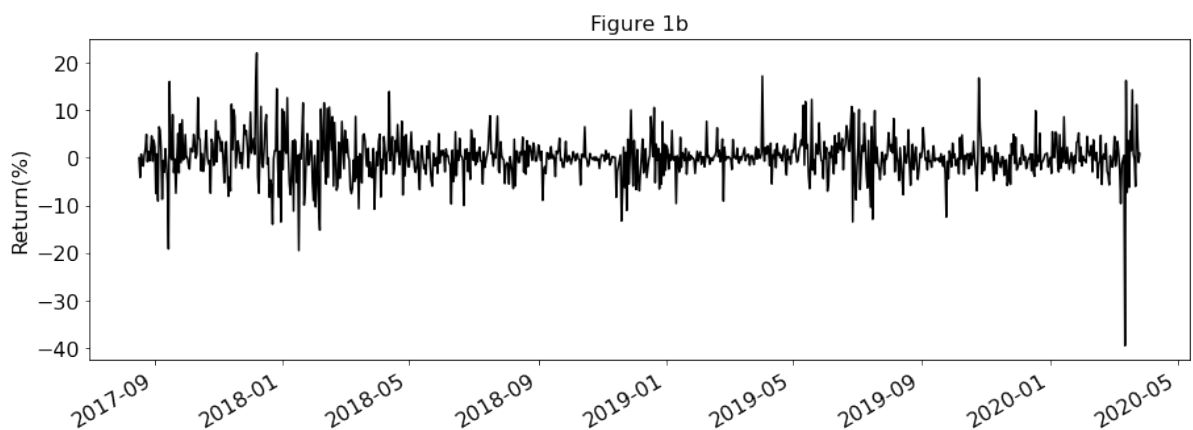

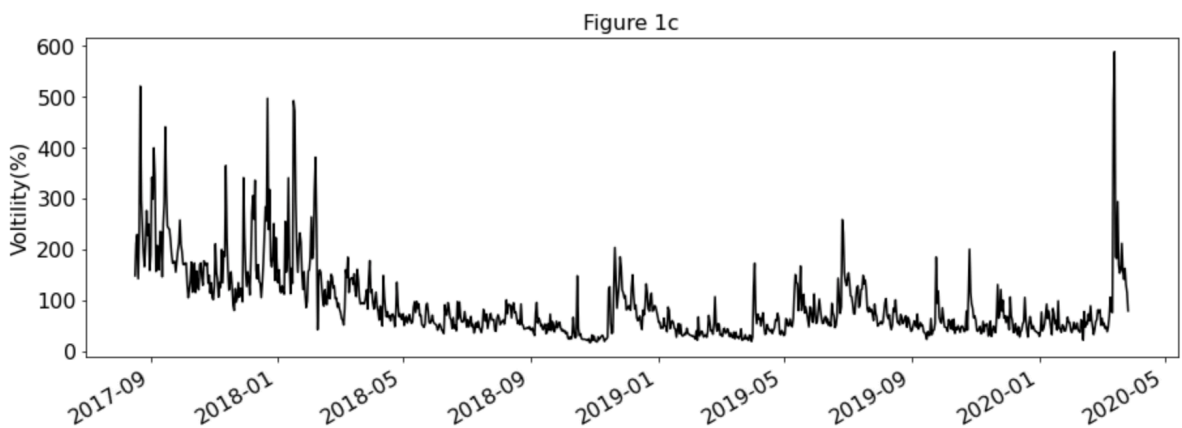

**Figure 2**
**GARCH model's volatility forecast vs. future realized volatility**

For each day *t* from August 17, 2018 onwards, we estimate the GARCH model with the whole historic daily data of BTC.USDT. We then generate the GARCH-model volatility forecast of day *t+1*. We also calculate the realized volatility of day *t+1* from the minute-level return data. Figure 2 plots the volatility forecast by the GARCH model and the future realized volatility, both of which are annualized. Our sample period is from August 17, 2017 to March 26, 2020. Our forecasting period is from August 17, 2018 to March 26, 2020.

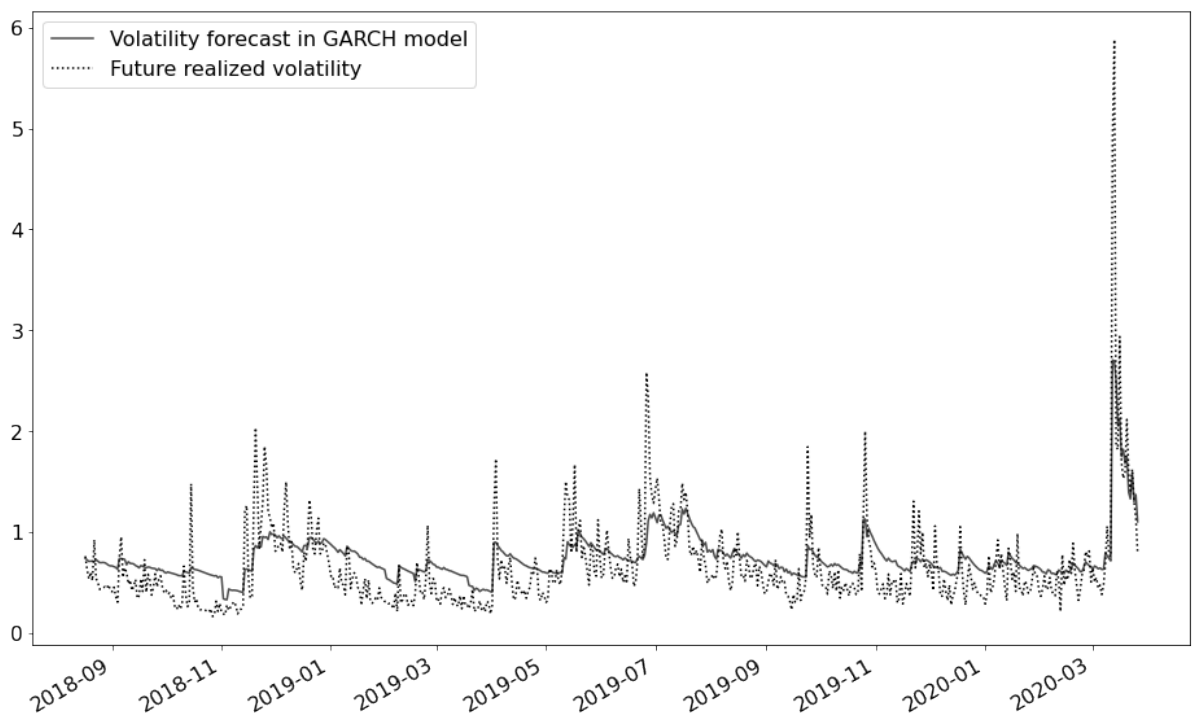

# Figure 3
## Various model's volatility forecast vs. future realized volatility

For each day *t* from August 17, 2018 onwards, we estimate the HIST, ARCH, EMA, and EGARCH models with the whole historic daily data of BTC.USDT. We then generate each model's volatility forecast of day *t+1*. We also calculate the realized volatility of day *t+1* from the minute-level return data. Figures 3a, 3b, 3c, and 3d plot the volatility forecast by each of the four models and the future realized volatility. Our sample period is from August 17, 2017 to March 26, 2020. Our forecasting period is from August 17, 2018 to March 26, 2020.

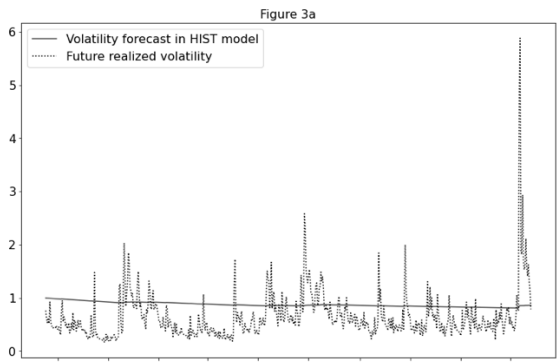
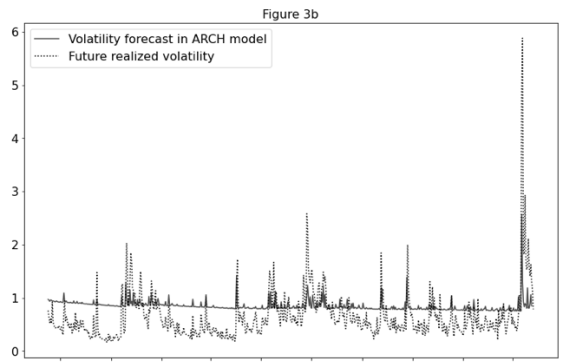
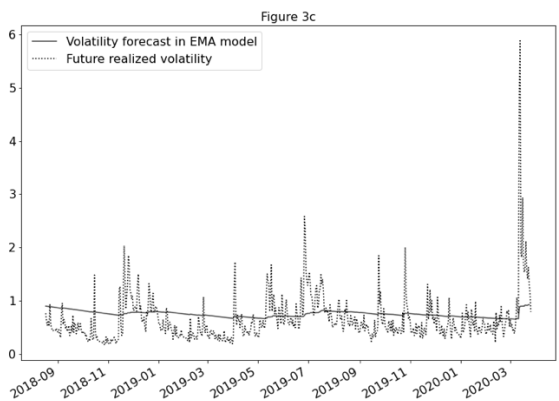
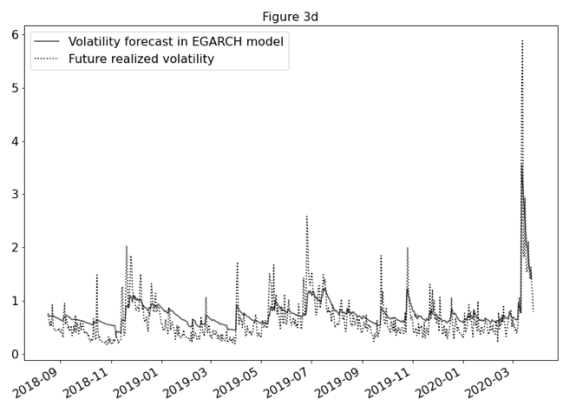

**Figure 4**
**GARCH model's multi-day average volatility forecast**

For each day *t* from September 10, 2019 onwards, we estimate the GARCH model with the whole historic daily data of BTC.USD index price quoted from Deribit. We then generate the GARCH-model volatility forecast of day *t+1, t+2,…,T*, where *T* marks the option's maturity date. That is, we are forecasting a daily volatility series all the way to option's maturity date. Next, we calculate the equal-weight average of this series to arrive at the GARCH-model's multi-day average volatility forecast. Additionally, we calculate day *t+1*'s volume-weighted average implied volatility of 27Mar2020 at-the-money call and put options. Figure 4 plots the single-day (*t+1*) volatility forecast by the GARCH model, the multi-day average volatility forecast by the GARCH model, and the option implied volatility. Our sample period is from September 10, 2019 to March 26, 2020.

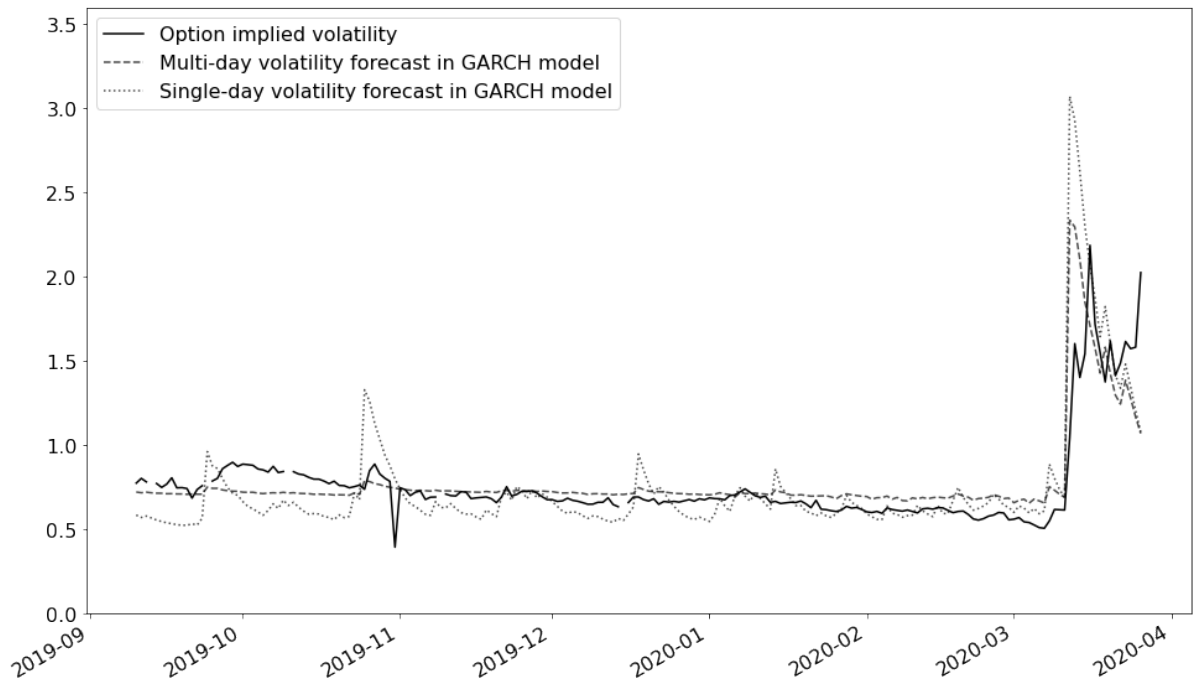

**Figure 5**
**Cumulative PNL of trading strategy**

Figure 5 shows the cumulative strategy PNL based on the volatility-spread trading strategy with parameters (24hour/0.05/0.00). Black dots mark the trading entries.

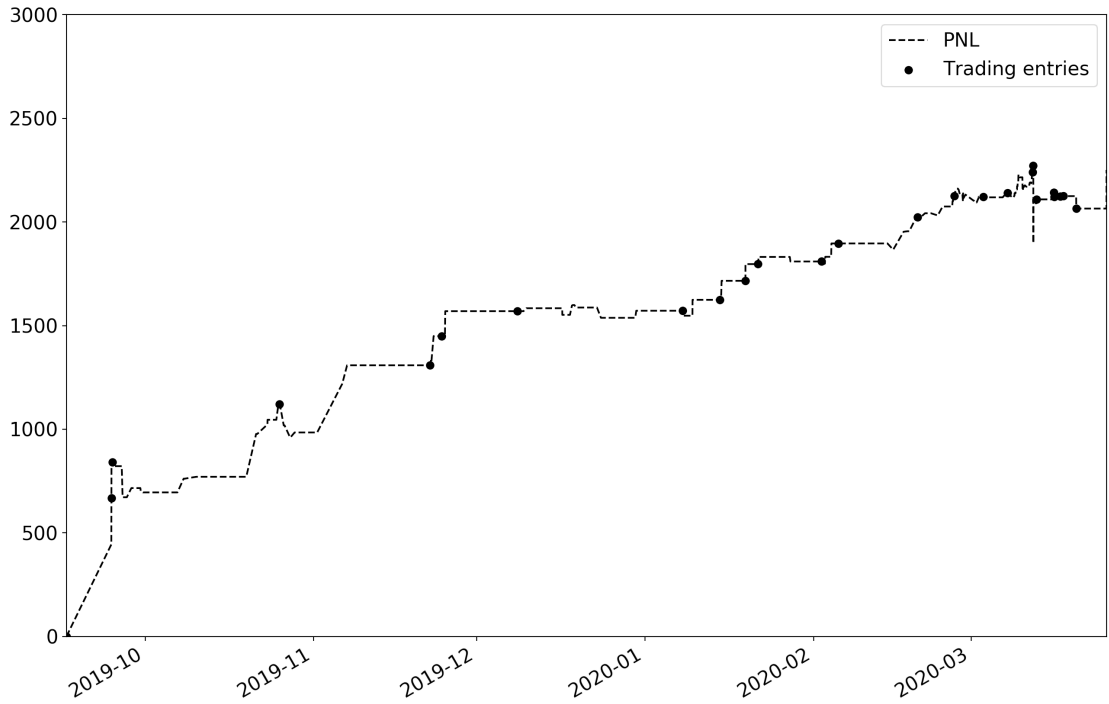

**Figure 6**
**Cumulative PNL of trading strategy incorporating volatility smile**

Figure 6 shows the cumulative strategy PNL based on the volatility-spread trading strategy with parameters (24hour/0.05/0.00), and the daily volatility forecast incorporating the volatility-smile adjustment. Black dots mark the trading entries.

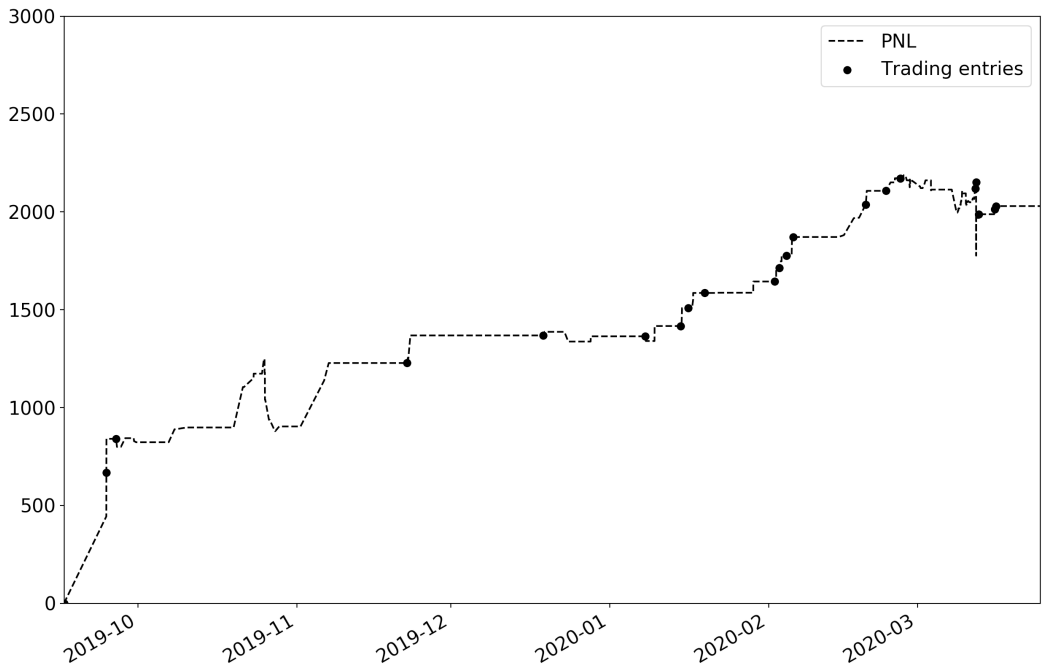

**Table 1**
**Statistical properties of BTC.USDT return series**

Table 1 shows the average return, annualized volatility, skewness, and excess kurtosis of BTC.USDT return series quoted from Binance at various (daily, 12-hour, 6-hour, 3-hour, and 1-hour) frequencies. Our sample period is from August 17, 2017 to March 26, 2020.

| Variable | Average return | Volatility (annualized) | Skewness | Kurtosis |
|---|---|---|---|---|
| Daily return | 0.15% | 85.83% | -0.52 | 8.66 |
| 12 hours return | 0.07% | 84.31% | 0.10 | 8.82 |
| 6 hours return | 0.04% | 86.16% | -0.19 | 11.21 |
| 3 hours return | 0.02% | 89.87% | 0.42 | 23.57 |
| 1 hour return | 0.01% | 97.23% | 0.03 | 29.74 |

**Table 2**

**In-sample-fit results for ARCH, GARCH and EGARCH models**

Table 2 reports the in-sample-fit outcomes for ARCH, GARCH, and EGARCH models. We report the model parameters' coefficients, and their associated t-statistics in parentheses. Additionally, we report each model's loglikelihood, AIC, and BIC values. *, **, and *** denote statistical significance at the levels of 10%, 5%, and 1% respectively. We use the BTC.USDT daily return series quoted from Binance for our model estimation. Our sample period is from August 17, 2017 to March 26, 2020, for a total of 953 daily observations.

| Variables | ARCH | GARCH | EGARCH |
|---|---|---|---|
| #Obs. | 953 | 953 | 953 |
| $a_0$ | 1.01e-03*** | 1.36e-04** | 1.60e-04** |
|  | (8.47) | (2.32) | (2.29) |
| $\alpha$ | 0.20*** | 0.11** | 0.08*** |
|  | (3.19) | (2.38) | (2.17) |
| $\beta$ |  | 0.83*** | 0.80*** |
|  |  | (17.45) | (13.80) |
| $\theta$ |  |  | 0.13 |
|  |  |  | (1.12) |
| loglikelihood | 1551.19 | 1663.62 | 1670.31 |
| AIC | -3096.39 | -3319.24 | -3330.62 |
| BIC | -3081.82 | -3299.79 | -3306.32 |

**Table 3**
**Distributional statistics for GARCH model's volatility forecast**

Table 3 reports the GARCH model's 1-day-ahead volatility forecast's various distributional statistics including the average, minimum, 5%, 10%, 25%, 50% (median), 75%, 90%, 95%, and maximum values of each variable. We report the distributional statistics of the GARCH model based on different look-back periods. *Forecast_vol* uses the look-back period of the whole period starting from August 17, 2017. *Forecast_vol365* uses the look-back period of the past 365 days. *Forecast_vol180* uses the look-back period of the past 180 days. *Forecast_vol90* uses the look-back period of the past 90 days. *Forecast_vol30* uses the look-back period of the past 30 days. Our sample period is from August 17, 2017 to March 26, 2020. Our forecasting period is from August 17, 2018 to March 26, 2020.

| Variable | Mean | Min | 5% | 10% | 25% | 50% median | 75% | 90% | 95% | Max |
|---|---|---|---|---|---|---|---|---|---|---|
| Forecast_vol | 71.5% | 32.9% | 49.4% | 57.4% | 61.7% | 68.4% | 79.9% | 92.5% | 100.1% | 123.4% |
| Forecast_vol365 | 67.2% | 39.7% | 48.5% | 53.9% | 58.1% | 64.0% | 73.2% | 87.1% | 94.2% | 135.3% |
| Forecast_vol180 | 65.4% | 33.9% | 46.3% | 49.5% | 56.4% | 63.2% | 70.4% | 83.3% | 88.8% | 179.8% |
| Forecast_vol90 | 65.7% | 26.8% | 39.8% | 43.7% | 52.2% | 61.3% | 75.7% | 89.9% | 100.2% | 210.3% |
| Forecast_vol30 | 61.8% | 16.4% | 27.0% | 33.7% | 42.6% | 56.6% | 74.6% | 95.9% | 111.7% | 234.0% |

**Table 4**
**Out-of-sample-forecast results for individual models**

We investigate the volatility-forecasting power of various models based on the following regression:

$$RV_{t+1} = \beta_0 + \beta_1 Fvol_t + e_{t+1}$$

where $RV_{t+1}$ is the future realized volatility of day $t+1$, calculated from minute-level BTC.USDT return series quoted from Binance; $Fvol_t$ is the model-specific volatility forecast for day $t+1$ based on data available by the end of day $t$. Models we consider for this exercise include the HIST, EMA, ARCH, GARCH, and EGARCH model. For each model, we also consider the following look-back windows: whole sample period, past 365 days, past 180 days, past 90 days, and past 30 days. Table 4 reports the regression coefficients, associated t-statistics in parentheses, adjusted R-square, and mean absolute error (MAE). *, **, and *** denote statistical significance at the levels of 10%, 5%, and 1% respectively. Our sample period is from August 17, 2017 to March 26, 2020. Our forecasting period is from August 17, 2018 to March 26, 2020.

|  | $\beta_0$ | $\beta_1$ | R^2 adjusted | MAE |
|---|---|---|---|---|
| HIST Past Whole | -1.11*** | -1.11*** | 0.98% | 28.28% |
|  | (-2.61) | (4.33) |  |  |
| HIST Past 365d | -0.04 | -0.04*** | -0.15% | 28.33% |
|  | (-0.27) | (5.15) |  |  |
| HIST Past 180d | 1.10*** | 1.10 | 3.60% | 27.99% |
|  | (4.85) | (-0.70) |  |  |
| HIST Past 90d | 0.91*** | 0.91 | 8.10% | 26.90% |
|  | (7.28) | (0.37) |  |  |
| HIST Past 30d | 0.86*** | 0.86* | 35.84% | 23.09% |
|  | (13.61) | (1.68) |  |  |
| EMA Past Whole | -0.45* | 1.47*** | 0.03% | 27.29% |
|  | (-1.87) | (4.55) |  |  |
| EMA Past 365d | -0.29 | 1.30*** | 0.04% | 27.86% |
|  | (-1.61) | (5.19) |  |  |
| EMA Past 180d | -0.40*** | 1.55*** | 15.02% | 25.44% |
|  | (-3.91) | (10.23) |  |  |
| EMA Past 90d | -0.14*** | 1.18*** | 22.93% | 23.06% |
|  | (-2.37) | (13.25) |  |  |
| EMA Past 30d | 0.06* | 0.88*** | 35.32% | 20.29% |
|  | (1.78) | (17.93) |  |  |

|                    | β0         | β1       | R^2 adjusted | MAE    |
|--------------------|------------|----------|--------------|--------|
| ARCH Past Whole    | -1.38***   | 2.38***  | 31.97%       | 24.60% |
|                    | (-11.27)   | (16.63)  |              |        |
| ARCH Past 365d     | -0.26***   | 1.23***  | 10.78%       | 27.42% |
|                    | (-2.46)    | (8.48)   |              |        |
| ARCH Past 180d     | -0.24***   | 1.31***  | 12.14%       | 25.71% |
|                    | (-2.47)    | (9.06)   |              |        |
| ARCH Past 90d      | 0.06       | 0.87***  | 11.52%       | 25.50% |
|                    | (1.01)     | (8.80)   |              |        |
| ARCH Past 30d      | 0.15***    | 0.74***  | 24.14%       | 22.31% |
|                    | (4.10)     | (13.70)  |              |        |
| GARCH Past Whole   | -0.29      | 1.26***  | 43.64%       | 19.51% |
|                    | (-6.36)    | (21.34)  |              |        |
| GARCH Past 365d    | -0.27***   | 1.31***  | 49.02%       | 18.53% |
|                    | (-6.68)    | (23.78)  |              |        |
| GARCH Past 180d    | 0.10**     | 0.78***  | 25.60%       | 21.57% |
|                    | (2.51)     | (14.24)  |              |        |
| GARCH Past 90d     | -0.01      | 0.95***  | 40.54%       | 20.25% |
|                    | (-0.33)    | (20.03)  |              |        |
| GARCH Past 30d     | 0.16***    | 0.72***  | 32.11%       | 20.46% |
|                    | (5.19)     | (16.69)  |              |        |
| EGARCH Past Whole  | -0.16***   | 1.07***  | 46.53%       | 18.93% |
|                    | (-4.32)    | (22.62)  |              |        |
| EGARCH Past 365d   | 0.00       | 0.91***  | 46.24%       | 19.92% |
|                    | (-0.04)    | (22.49)  |              |        |
| EGARCH Past 180d   | 0.15***    | 0.71***  | 41.49%       | 21.40% |
|                    | (5.39)     | (20.42)  |              |        |
| EGARCH Past 90d    | 0.18***    | 0.67***  | 40.65%       | 20.98% |
|                    | (6.71)     | (20.07)  |              |        |
| EGARCH Past 30d    | 0.20***    | 0.69***  | 32.78%       | 20.29% |
|                    | (6.71)     | (16.95)  |              |        |

**Table 5**
**Summary statistics on Deribit BTC option contracts**

Table 5 reports the quarterly summary statistics regarding the BTC option contracts traded on the Deribit exchange. Specifically, we report the number of traded contracts, the total number of trades, and the total trading volume in USD. The trading volume is calculated based on the option premium in USD. Our sample period is from October 1, 2018 to March 31, 2020.

| Quarter | # Contracts | # Trades | $ Trading Volume |
|---------|-------------|----------|------------------|
| 4Q2018  | 236         | 34,587   | 382,276,866      |
| 1Q2019  | 253         | 37,135   | 485,619,294      |
| 2Q2019  | 387         | 64,811   | 1,453,655,901    |
| 3Q2019  | 348         | 67,475   | 1,762,728,341    |
| 4Q2019  | 377         | 73,593   | 1,667,613,979    |
| 1Q2020  | 971         | 121,041  | 3,334,233,932    |

**Table 6**
**Out-of-sample-forecast results for various model combinations**

We investigate the volatility-forecasting power of various model combinations based on the following regression:

$$RV_{t+1} = \beta_0 + \beta_1 * ARCH_t + \beta_2 * GARCH_t + \beta_3 * EGARCH_t + \beta_4 * IV_t + e_{t+1}$$

where $RV_{t+1}$ is the future realized volatility of day $t+1$, calculated from minute-level BTC.USD index price quoted from Deribit; ARCH, GARCH, and EGARCH 1-day-ahead volatility forecasts are defined in the same manner as described in Table 4, with the whole sample period up to day $t$ as the look-back window; IV is defined as the volume-weighted average implied volatility of at-the-money call and put options on day $t$. Table 6 reports the regression coefficients, associated t-statistics in parentheses, and adjusted R-squares of various regression models' outcomes. ∗, ∗∗, and ∗∗∗ denote statistical significance at the levels of 10%, 5%, and 1% respectively. Our sample period is from August 17, 2018 to March 26, 2020. Our forecasting period is from September 10, 2019 to March 26, 2020.

| β0 | β1 | β2 | β3 | β4 | R^2 adjusted |
|---|---|---|---|---|---|
| -2.74*** | 4.38*** | | | | 28.93% |
| (-7.78) | (8.77) | | | | |
| -0.83*** | | 1.61*** | | | 63.83% |
| (-12.03) | | (18.26) | | | |
| -0.68*** | | | 1.40*** | | 64.92% |
| (-11.34) | | | (18.70) | | |
| -0.68*** | | | | 1.35*** | 31.83% |
| (-5.99) | | | | (9.39) | |
| -0.7*** | | 1.93*** | | -0.48** | 64.94% |
| (-8.55) | | (13.47) | | (-2.83) | |
| -0.63*** | | | 1.48*** | -0.15 | 64.73% |
| (-7.66) | | | (13.39) | (-0.99) | |
| -2.61*** | 3.11*** | | | 1.01*** | 44.43% |
| (-8.35) | (6.52) | | | (7.24) | |

**Table 7**
**Trading performance of the volatility-spread strategy**

The first three columns specify the volatility-spread trading strategy parameters: GARCH model's sampling frequency, the entry threshold of volatility spread, and the exit threshold of volatility spread. The last five columns show the number of trades, win/loss ratio, win rate, total PNL, and average PNL of the volatility-spread trading strategies. Our sample period is from September 10, 2019 to March 26, 2020.

| GARCH sampling frequency | Entry threshold | Exit threshold | # trades | Win/loss ratio | Win rate | Total PNL ($) | PNL per trade ($) |
|---|---|---|---|---|---|---|---|
| 24-hour | 0.05 | 0.00 | 25 | 10.09 | 88.5% | 2,251 | 90 |
| 24-hour | 0.075 | 0.05 | 10 | 10.53 | 90.0% | 1,553 | 155 |
| 24-hour | 0.075 | 0.00 | 18 | 10.80 | 89.5% | 2,275 | 126 |
| 24-hour | 0.10 | 0.05 | 8 | 10.93 | 87.5% | 1,478 | 185 |
| 24-hour | 0.10 | 0.00 | 13 | 12.28 | 92.3% | 2,233 | 172 |

## Table 8
## Trade summary (24-hour/0.05/0.00)

This table reports a detailed trade-level summary for the 24-hour GARCH model with entry threshold at 0.05 and exit threshold at 0.00. For each trade, we report the timestamp of each trade entry and exit, the volatility spread at the time of each entry and exit, direction of the trade of the call option, option premium at the time of each entry and exit, the total PNL of each round-trip trade, and the PNL that belongs to the trading of the underlying Bitcoin perpetual swap, and the PNL that belongs to the trading of the call option. Our sample period is from September 10, 2019 to March 26, 2020.

| Timestamp | Entry | Exit | Direction | Option premium ($) | PNL total ($) | PNL underlying ($) | PNL option ($) |
|---|---|---|---|---|---|---|---|
| 2019-09-16 12:40 | -0.23 | | sell to open | 3,770 | | | |
| 2019-09-24 19:10 | 0.05 | 0.05 | buy to close buy to open | 2,007 | 667 | -1,096 | 1,764 |
| 2019-09-24 19:17 | | -0.03 | sell to close | 2,125 | 173 | 55 | 118 |
| 2019-09-24 23:56 | -0.07 | | sell to open | 2,161 | | | |
| 2019-10-25 12:42 | | 0.00 | buy to close | 1,301 | 280 | -580 | 860 |
| 2019-10-25 17:21 | -0.07 | | sell to open | 2,149 | | | |
| 2019-11-07 4:02 | | 0.02 | buy to close | 2,323 | 188 | 362 | -174 |
| 2019-11-22 10:04 | -0.06 | | sell to open | 1,037 | | | |
| 2019-11-23 2:12 | | 0.03 | buy to close | 943 | 141 | 47 | 94 |
| 2019-11-24 14:12 | 0.05 | | buy to open | 852 | | | |
| 2019-11-25 5:36 | | -0.01 | sell to close | 676 | 120 | 296 | -176 |
| 2019-12-08 11:18 | 0.05 | | buy to open | 950 | | | |
| 2019-12-30 9:11 | | 0.00 | sell to close | 805 | 2 | 147 | -145 |
| 2020-01-07 22:21 | -0.06 | | sell to open | 1,146 | | | |
| 2020-01-09 17:38 | | 0.00 | buy to close | 963 | 53 | -130 | 182 |
| 2020-01-14 18:38 | -0.09 | | sell to open | 1,610 | | | |
| 2020-01-15 1:55 | | 0.01 | buy to close | 1,537 | 92 | 18 | 74 |
| 2020-01-19 11:24 | -0.05 | | sell to open | 1,471 | | | |
| 2020-01-19 11:25 | | 0.01 | buy to close | 1,383 | 81 | -8 | 88 |
| 2020-01-21 17:55 | 0.06 | | buy to open | 1,291 | | | |
| 2020-01-27 18:06 | | 0.00 | sell to close | 1,390 | 13 | -86 | 98 |
| 2020-02-02 10:59 | -0.05 | | sell to open | 1,898 | | | |
| 2020-02-04 7:01 | | 0.01 | buy to close | 1,679 | 87 | -132 | 219 |
| 2020-02-05 14:25 | -0.06 | | sell to open | 1,864 | | | |
| 2020-02-19 22:07 | | 0.01 | buy to close | 1,820 | 127 | 83 | 44 |
| 2020-02-20 2:47 | -0.06 | | sell to open | 1,867 | | | |
| 2020-02-26 16:31 | | 0.01 | buy to close | 1,214 | 103 | -551 | 654 |

| Timestamp | | Entry | Exit | Direction | Option premium ($) | PNL total ($) | PNL underlying ($) | PNL option ($) |
|---|---|---|---|---|---|---|---|---|
| 2020-02-26 | 22:03 | 0.05 | | buy to open | 1,046 | | | |
| 2020-03-02 | 9:32 | | 0.00 | close to close | 989 | -4 | 0 | 4 |
| 2020-03-03 | 5:24 | 0.05 | | buy to open | 1,064 | | | |
| 2020-03-07 | 11:15 | | 0.00 | sell to close | 1,275 | 17 | -194 | 211 |
| 2020-03-07 | 16:37 | 0.06 | | buy to open | 1,282 | | | |
| 2020-03-12 | 4:33 | | 0.00 | sell to close | 266 | 101 | 1,118 | -1,017 |
| 2020-03-12 | 7:14 | -0.05 | | sell to open | 218 | | | |
| 2020-03-12 | 7:39 | | 0.01 | buy to close | 192 | 31 | 6 | 25 |
| 2020-03-12 | 10:16 | -0.05 | | sell to open | 186 | | | |
| 2020-03-13 | 0:00 | 0.66 | 0.66 | sell to close buy to open | 64 | -163 | -285 | 122 |
| 2020-03-16 | 1:48 | | -0.01 | sell to close | 98 | 34 | 0 | 34 |
| 2020-03-16 | 7:51 | -0.06 | | sell to open | 53 | -20 | 0 | -20 |
| 2020-03-17 | 8:49 | | 0.02 | buy to close | 53 | 0 | 0 | 0 |
| 2020-03-17 | 9:09 | 0.05 | | buy to open | 51 | | | |
| 2020-03-17 | 12:31 | | -0.01 | sell to close | 53 | 2 | 0 | 2 |
| 2020-03-18 | 0:11 | -0.11 | | sell to open | 48 | | | |
| 2020-03-20 | 8:07 | | 0.02 | buy to close | 108 | -60 | 0 | -60 |
| 2020-03-20 | 8:20 | -0.08 | | sell to open | 201 | | | |

**Table 9**
**Trading performance incorporating volatility smile**

The first three columns specify the volatility-smile-adjusted volatility-spread trading strategy parameters: GARCH model's sampling frequency, the entry threshold of volatility spread, and the exit threshold of volatility spread. The last five columns show the number of trades, win/loss ratio, win rate, total PNL, and average PNL of the volatility-spread trading strategies. Our sample period is from September 10, 2019 to March 26, 2020.

| GARCH sampling frequency | Entry threshold | Exit threshold | # trades | Win/loss ratio | Win rate | Total PNL ($) | PNL per trade ($) |
|---|---|---|---|---|---|---|---|
| 24-hour | 0.050 | 0.00 | 21 | 10.18 | 95.2% | 2,028 | 97 |
| 24-hour | 0.075 | 0.05 | 12 | n/a | 100.0% | 2,178 | 182 |
| 24-hour | 0.075 | 0.00 | 14 | 10.02 | 92.9% | 1,692 | 121 |
| 24-hour | 0.100 | 0.05 | 7 | 10.34 | 75.0% | 1,299 | 186 |
| 24-hour | 0.100 | 0.00 | 7 | 8.93 | 85.7% | 1,151 | 164 |